\documentstyle [12pt] {article}
\def\lsim{\lower.5ex\hbox{$\; \buildrel < \over \sim \;$}}
\def\gsim{\lower.5ex\hbox{$\; \buildrel > \over \sim \;$}} 
\def \simeq{\lower.3ex\hbox{$\; \buildrel \sim \over - \;$}}
\begin{document}
\parskip 14pt
\parindent 12pt
\noindent {\bf \large Aspects of Accretion Processes On a Rotating Black Hole}\\

\noindent {Sandip K. Chakrabarti}\\
\noindent {\small Tata Institute of Fundamental Research, Homi Bhabha Road, 
Mumbai, 400005}
\pagenumbering{arabic}

\noindent {\small {\bf Abstract:} We describe 
the most general nature of accretion and wind flows
around a compact object and emphasize on the properties
which are special to {\it black hole} accretion.
The angular momentum distribution in the most 
general solution is far from Keplerian, and the non-Keplerian 
disks can include standing shock waves. We also present 
fully time dependent numerical simulation results to show
that they agree with these analytical solutions.
We describe the spectral properties of these accretion disks 
and show that the soft and hard states of the black hole candidates
could be explained by the change of the accretion rate of the disk.
We present fits of the observational data to demonstrate the presence of
sub-Keplerian flows around black holes.}

\noindent {e-mail: chakraba@tifrc2.tifr.res.in}\\

\noindent Presented as a Plenary talk at the 17th Conference of
Indian Association of General Relativity and Gravitation.

\baselineskip 12pt

\noindent {\bf Introduction}

Centers of galaxies are believed to be the seats of 
massive black holes ($M_{BH}  \sim 10^{6-9} M_\odot$) and some evolved
compact binary systems are believed to be the seats of small mass black
holes ($M_{BH} \sim 3-10 M_\odot$). In the 1970s, the standard 
accretion disk models  around black holes were constructed
(Shakura \& Sunyaev 1973; Novikov \& Thorne 1973)
assuming Keplerian angular momentum distribution in the orbiting matter.
In the early 1980s, these disks became the favorite candidates for 
the explanation of the ``big blue bump'' seen in the UV region of the
continuum spectra of active galaxies (Malkan 1982; Sun \& Malkan 1989)
with some success.

However, from X-ray and $\gamma$-ray spectra and line emissions 
from these objects (see, Chakrabarti, 1996a for a general review), it is 
becoming clear that the accretion disks cannot be simple Keplerian type
nor are as simple as spherically symmetric Bondi 
flows (Bondi, 1952). To satisfy the inner boundary condition on the horizon,
matter must cross the horizon supersonically (Chakrabarti 1990, hereafter C90)
and therefore must pass through a sonic point where the
radial velocity of the flow is the same as the sound velocity. This means
that the angular momentum must deviate from a Keplerian flow. 
All these considerations require one to solve the most general set of 
equations which must include  effects of rotation, flow pressure, 
viscosity, advection, heating and cooling processes. Below, we 
present these equations and discuss the solutions and their
implications. For the purpose of the present review we emphasize
those properties which are {\it special} to a black hole accretion.

We consider a perfect flow around a Kerr black hole with the metric
(e.g., Novikov \& Thorne, 1973; using $t$, $r$, $\theta$ \& $z$ as coordinates
and the units which render $G=M_{BH}=c=1$).
$$
ds^2 = -\frac{r^2 \Delta}{A} dt^2 + \frac{A}{r^2} 
(d\phi-\omega dt)^2 +\frac{r^2}{\Delta} dr^2 + dz^2,
\eqno{(1)}
$$
where, 
$$
A= r^4 + r^2 a^2 + 2 r a^2,
$$
$$
\Delta= r^2 - 2 r + a^2,
$$
$$
\omega = \frac{2 a r}{A}.
$$
Here, $a$ is the Kerr parameter.

The stress-energy tensor of a perfect fluid with pressure $p$ and mass density
$\rho=\rho_0(1+\pi)$, $\pi$ is the internal energy is give by,
$$
T_{\mu\nu} = \rho u_\mu u_\nu + p(g_{\mu\nu} + u_\mu u_\nu).
\eqno{(2)}
$$
We shall concentrate on the time independent solution of the governing
equations (Chakrabarti 1996b, hereafter C96b).
The equation for the balance of the radial momentum is obtained from 
$(u_\mu u_\nu + g_{\mu\nu}) T^{\mu\nu}_{;\nu}=0$.
Here the advection term due to significant radial velocity is included.
The baryon number conservation equation (continuity equation) 
is obtained from $(\rho_0 u^\mu)_{; \mu}=0$. The conservation of 
angular momentum is obtained from $(\delta_\mu^\phi T^{\mu \nu})_{;\nu}=0$. 
Here the angular momentum is allowed to be non-Keplerian.
Entropy generation equation is obtained from the first law of thermodynamics
along with baryon conservation equation: $(S^\mu)_{;\mu}=\frac{1}{T}
[2\eta \sigma_{\mu\nu}\sigma^{\mu\nu}]$, where, $T$ is the temperature 
of the flow and $\eta$ is the coefficient of viscosity. $\sigma_{\mu\nu}$
is the shear tensor which is responsible for the viscous transport
of angular momentum.

These set of equations are solved simultaneously
along with the possibility that the shock waves may form in the flow, 
where, the following Rankine-Hugoniot conditions must be fulfilled:
$$
W_-n^\nu + (W_-+\Sigma_-) (u_-^\mu n_\mu)u_-^\nu  = 
W_+n^\nu + (W_++\Sigma_+) (u_+^\mu n_\mu) u_+^\nu .
$$
Here, $n_\mu$ is the four normal vector component across the shock, 
and $W$ and $\Sigma$ are vertically integrated pressure and density on 
the shock surface.

\noindent {\bf Solution Topologies:}

First note that the above mentioned equations are valid for any compact 
object whose external spacetime is similar to that of a Kerr black hole.
However, on a neutron star surface matter has to stop and
corotate with the surface velocity. The inner boundary condition is therefore
sub-sonic. On a black hole, on the other hand, the flow must enter through the
horizon with the velocity of light, and therefore must be supersonic.
Stationary shock waves may form when matter starts piling up behind the
centrifugal barrier (which arises due to centrifugal force $\propto
\lambda^2/r^3$). The supersonic flow becomes subsonic at the shock
and again becomes supersonic before entering through the horizon. 
Clearly, the flow has to become supersonic, before forming the shock 
as well, and therefore pass through another sonic point
at a large distance away from the black hole. Thus, as a whole,
the flow may deviate from a Keplerian disk and  (a) enter through the
inner sonic point only, or, (b) enter through the outer sonic point only, or, 
(c) pass through the outer sonic point, then a shock, and finally through an
inner sonic point if the shock conditions are satisfied.
If the angular momentum is too small, then the flow has only one sonic point
and shocks cannot be formed as in a Bondi flow. For more details, see 
Chakrabarti (1989, hereafter C89), C90 and C96a.

Three flow parameters govern the topology of the flow: the $\alpha$ parameter
(Shakura \& Sunyaev, 1973) which determines the viscosity, 
the location of the inner sonic point $r_{in}$ through which matter must pass 
through, and the specific angular momentum $\lambda_{in}$ of the matter at 
the horizon  (or, that at $r_{in}$). It so happens that these parameters 
are sufficient to completely determine the solution when a cooling process
is provided. To illustrate the flow topologies, we first choose the
viscosity of the flow to be negligible (C96b). In the inviscid case, the 
angular momentum and energy remain constant $hu_\phi= l$ and $hu_t={\cal E}$.
Hereafter, we use $\lambda=l/{\cal E}$ to be the specific angular
momentum. Flow entropy remains constant unless it passes through a
shock wave where it goes up within a thin layer. In Fig. 1 we show 
{\it all possible} solutions of weakly viscous accretion flows around 
a Kerr black hole of rotation parameter $a=0.5$. 
(For flows in Schwarzschild geometry, see, C89, C90.)
The adiabatic index $\gamma=4/3$ has been chosen.
Vertical equilibrium flow model with the vertical
height prescription of NT73 is used. In the central box,
we divide the parameter space spanned by ($\lambda, {\cal E}$)
into nine regions marked by $N$, $O$, $NSA$, $SA$, $SW$, $NSW$, $I$,
$O^*$, $I^*$. The horizontal line at ${\cal E}=1$ corresponds to the rest 
mass of the flow. Surrounding this parameter space, we plot various
solution topologies (Mach number $M=v_r/a_s$ vs. logarithmic radial distance
where $v_r$ is the radial velocity and $a_s$ is the sound speed) marked 
with the same notations (except $N$). Each of these solution topologies has 
been drawn using flow parameters from the respective region of the central
box. Though each contour of each of the boxes represents individual
solutions (differing only by specific entropy), 
the relevant solutions are the ones which are self-crossing,
as they are transonic. The crossing points are `X' type or saddle
type sonic points and the contours of circular topology surround `O' type 
sonic points. If there are two `X' type sonic points, the inner
one is called the inner sonic point, and the outer one is called the
outer sonic point. If there is only one `X' type sonic point
in the entire solution, then the terminology of inner or outer
is used according to whether the sonic point is close to or away from the
black hole. The solutions from the region `O' has only the outer sonic point.
The solutions from the regions $NSA$ and $SA$ have two `X' type sonic points
with the entropy density $S_o$ at the outer sonic point {\it less} than the
entropy density $S_i$ at the inner sonic point. However, flows from $SA$
pass through a standing shock (See Fig. 2) as the Rankine-Hugoniot
condition is satisfied. The entropy generated at the shock is exactly 
$S_i-S_o$, which is advected towards the hole to enable the flow to pass 
through the inner sonic point. Rankine-Hugoniot condition is not satisfied
for flows from the region $NSA$. Numerical simulation indicates
(Ryu, Chakrabarti \& Molteni, 1996) that the solution is very unstable
and show periodic changes in emission properties as the flow
constantly try to form the shock wave, but fail to do so. The solutions
from the region $SW$ and $NSW$ are very similar to  those from 
$SA$ and $NSA$. However, $S_o \geq S_i$ in these cases.
Shocks can form only in winds from the region $SW$. The shock condition is not
satisfied in winds from the region $NSW$. This may make the $NSW$ flow 
unstable as well. A flow from region $I$ only has the inner sonic 
point and thus can form shocks (which require 
the presence of two saddle type sonic points) only if the inflow
is already supersonic due to some other physical processes.

The flows from regions $I^*$ and $O^*$ are interesting in the sense
that each of them has two sonic points (one `X' and one `O')
only and neither produces complete and global solution. The region $I^*$ 
has an inner sonic point but the solution does not extend subsonically
to a large distance. The region $O^*$ has an outer sonic point, but the
solution does not extend supersonically 
to the horizon! In both the cases a weakly viscous
flow is expected to be unstable. When a significant viscosity is added, the closed
topology of $I^*$ is expected to open up (Fig. 3 below; Chakrabarti, 1996c;
hereafter C96c) and then the flow can join with a Keplerian disk.  

\noindent {\bf Examples of `Discontinuous' Solutions: }

In Fig. 1 we presented continuous solutions and mentioned the possibility
of shock formation.
In Fig. 2(a-b) we give examples of solutions which include shock wave
discontinuities. Mach number is plotted along Y-axis and logarithmic radial
distance is plotted along X-axis. Fig. 2a is drawn with parameters from the 
region $SA$ and Fig. 2b is drawn with parameters from the region $SW$.

In Fig. 2a, the single arrowed curve represents a solution 
coming subsonically from a large distance and becoming supersonic
at O, the outer sonic point. Subsequently, the flow jumps onto the
subsonic branch (at the place where Rankine-Hugoniot condition
is satisfied) along the vertical dashed line and subsequently the
flow enters the black hole through the inner sonic point at I
(double arrowed curve). The flow chooses to have a shock 
since the inner sonic point has a higher entropy. The 
parameters of the flow are $a=0.5$, ${\cal E}=1.003$ and $\lambda=3$.
In Fig. 2b, on the other hand, the accretion flow can straightaway pass 
through the inner sonic point (single arrowed curve) and will have no shocks.
However, outgoing winds, which may be originated closer to the black hole
can have a shock discontinuity (double arrowed curves). Here the flow
first passes through $I$, the shock (vertical dashed line) and the outer
sonic point `O'. The parameters in the case are $a=0.5$, ${\cal E}=1.007$
and $\lambda=3$. 

Above mentioned discussions are valid for flows around a black hole
only. For flows around a neutron star, the inner boundary condition
must be subsonic, and therefore, the shock transition (in Fig. 2a,
for example) must take the flow to a branch which will remain
subsonic thereafter, instead of a branch which is likely 
to become supersonic again.

\noindent{\bf Properties of Viscous Transonic Flows}

The topologies shown above become more complicated once various cooling
and viscous heating effects are included (C90, C96c).
We present the case of `isothermal gas' where the flow adjusts heating 
and cooling in such a way that matter remains at a constant 
temperature. Figs. 3(a-d) show the `phase space' of the accretion flow. 
We assume Schwarzschild black hole $a=0$ for simplicity and consider 
Shakura-Sunyaev (1973) viscosity prescription where the
viscous stress at a given location depends on the local thermal 
pressure $p$: $t_{r\phi}=-\alpha p$. 
We note the general change in topology of the flow. First of all, the 
circular topology of the inviscid flow (`O' type sonic point)
is converted into spiral topology as in a damped harmonic oscillator (C90). 
Secondly, the closed topology has opened up, partially or completely,
depending upon flow parameters. In Fig. 3a, the flow parameters are 
$\alpha=0.05$, $\lambda_{in}=1.8$, $r_{in}=2.8$.
The spiral is `half closed'. This topology is still good for shock formation
as shown in the vertical curve (C90). In Fig. 3b, we use, $\alpha=0.1$, 
$\lambda_{in}=1.8$, $r_{in}=2.8$. Here the spiral is complete with branches
from both sonic points. In Fig. 3c, we use, $\alpha=0.05$, 
$\lambda_{in}=1.77$, $r_{in}=2.8$ 
and in Fig. 3d, we use, $\alpha=0.05$, $\lambda_{in}=1.8$, $r_{in}=2.65$. 
The topology remains similar to Fig. 3b, but whereas Fig. 3b is attained by
increasing viscosity, Fig. 3c and Fig. 3d are obtained by decreasing angular
momentum and decreasing the inner sonic point respectively. What is common
in Fig. 3(b-d), is that both the sonic points allow flows to become Keplerian
(where, $M \sim 0$) but since the dissipation in the flow
passing through `I' is higher, we believe that the disk will choose this 
branch. More importantly, the flow cannot have a standing shock wave 
with this topology. All these solutions, except where $M\sim 0$, are 
sub-Keplerian as illustrated in Figs. 3(e-h) where the flow
angular momentum (solid curve) is compared with Keplerian
angular momentum (dotted curve). The location where the flow joins 
a Keplerian disk may be somewhat turbulent (description of which 
is not within the scope of our solution) so as to adjust pressures
of the subsonic Keplerian disk and the sub-Keplerian transonic viscous flow.
These figures illustrate the existence of critical
viscosity parameter $\alpha_c$, critical angular momentum $\lambda_c$,
and critical inner sonic point $r_c$ at which the flow
topologies are changed (C90).

The above results obtained for isothermal flow can be generalized easily
using the most general set of equations with heating and cooling
processes (C96c).  If $Q^+$ and $Q^-$ denote the heating and
cooling rates, and if, for illustration purpose, one assumes
that $f=(Q^+-Q^-)/Q^-= constant$, then one can easily solve the general 
equations to find the degree at which the flow deviates from
a Keplerian disk. In Fig. 4a we show the ratio $\lambda_{disk}/
\lambda_{Kep}$ for various viscosity and cooling parameters $\alpha$
and $f$. Clearly as the flow starts deviating from a Keplerian
disk $\lambda_{disk}/\lambda_{Kep}=1$, it becomes about 20-30\%
of Keplerian. As it approaches the black hole, it becomes close to
Keplerian again (and sometimes super-Keplerian also) before
plunging in to the black hole in a very sub-Keplerian manner.
In Fig. 4b, we show the ratio $v_r/v_\phi$ for the same disks. Note that
near the outer edge, as the flow deviates from a Keplerian disk,
$v_r <<v_\phi$, i.e., the flow is rotation dominated. Around $r\sim
100$ the radial velocity becomes dominant, and subsequently, even closer
to the black hole, the flow becomes rotation dominated (due to
the centrifugal barrier). Near the horizon the radial (advection)
velocity dominates once more. The case $f=0.5$, $\alpha=0.05$
showing a sudden jump in velocity ratio represents a solution
which includes a standing shock. The corresponding
angular momentum distribution in the upper panel does not
show discontinuity since shear is chosen to be continuous
across a shock wave. The region where the disk
deviates from a Keplerian disk could be geometrically
thick and would produce thick accretion disks. The post-shock
region, where the flow suddenly becomes hot and puffed up would
also form another thick disk of smaller size.

The solutions presented so far is of most general nature encompassing
the entire parameter space. Any other solutions of black hole
accretion or winds are special cases of these solutions. 

\noindent{\bf Problem of Angular Momentum Transport}

A curious property of a black hole accretion flow is that the shear stress
$\sigma_{r\phi}$ is not a monotonic function of distance, and it can be
negative close to the black hole. In the Newtonian geometry, it has the 
form $ - r \frac{d\Omega}{dr}$ and its magnitude is monotonically increasing 
inward. For a cold radial flow below the marginally bound orbit this
was pointed out by Anderson \& Lemos (1988). In Fig. 5(a-b) we show this for 
the complete and exact global solutions of the accretion flow presented
in the earlier Section (C96b). In Fig. 5a,
we present the variation of shear and angular velocity gradient for a 
typical prograde flow ($a=0.95$, $\lambda=2.3$ and ${\cal E}=1.001$) and in 
Fig. 5b we show the results of a retrograde flow ($a=0.95$, $\lambda=-4.0$ 
and ${\cal E}=1.0025$). The solid curves are computed using the most general
definition of shear tensor: $\sigma_{\alpha\beta}=
1/2 (u_{\alpha;\mu} P^\mu_\beta+u_{\beta;\mu}P^\mu_\alpha)- \Theta 
P_{\alpha\beta}/3$ where $P_{\alpha\beta}= g_{\alpha \beta} +u_\alpha u_\beta$ 
is the projection tensor and $\Theta=u^\alpha_{;\alpha}$ is the expansion.
The velocity components have been taken from the exact solution which 
passes through the outer sonic point. The subscript $+$ sign under
$\sigma^r_\phi$ refers to the supersonic branch of the solution.
We also present the same quantity (long dashed curve)
for the subsonic branch ($\sigma^r_{\phi-}$)
which passes through the outer sonic point. For comparison, we present
the component of shear tensor $\sigma^r_\phi|_{rot}$ (short dashed curve)
where we ignore the radial velocity completely $v_r <<v_\phi$.  We also 
present the variation of $d\Omega/dr$ (dotted curve) to indicate the
relation between the angular velocity gradient and shear.

Some interesting observations emerge from this peculiar shear distribution.
First, for prograde flows, the shear reverses its sign and becomes
negative just outside the horizon. This shows that very close to the horizon,
the angular momentum transport takes place 
`towards the black hole' rather than away from it,
Second, $\sigma^r_\phi|_{rot}$ and  $\sigma^r_{\phi-}$ which have
negligible radial velocities, vanish on the horizon whereas $\sigma^r_{\phi+}$ 
does not. Existence of the negative shear component 
shows the $\alpha$ viscosity prescription (SS73) is invalid close to 
a horizon, since pressure cannot be negative. However, the error by choosing 
a positive (e.g., $-\alpha p$) shear may not be significant, since one 
requires a (unphysically) large viscosity so as to transport significant 
angular momentum inwards. In any case, the inward angular momentum transport
 may change angular momentum distribution near the horizon (from that of 
an almost constant to something perhaps increasing inwards, similar
to the Keplerian distribution below marginally stable orbit).
Third, in retrograde flows, the $\sigma^r_{\phi+}$ reverses twice but
$\sigma^r_{\phi-}$ and $\sigma^r_\phi|_{rot}$ reverse once 
since they vanish on the horizon as in the prograde flows. 

The physical mechanism underlying the reversal of 
the shear component is simple. In general relativity, all the energies couple
one another. It is well known that the `pit in the potential' 
of a black hole is due to coupling between the rotational energy and 
and the gravitational energy (Chakrabarti, 1993).
As matter approaches the black hole, the rotational
energy, and therefore `mass' due to the energy increases which is also
attracted by the black hole. This makes gravity
much stronger than that of a Newtonian star. When the black hole
rotates, there are more coupling terms (such as thise arise out of
spin of the black hole and the orbital angular momentum of the matter)
which either favour gravity or go against it depending on whether
the flow is retrograde or prograde (Chakrabarti \& Khanna, 1992) respectively.
This is the basic reason why retrograde and prograde flows
display different reversal properties.

\noindent {\bf Numerical Simulations of Black Hole Accretion}

The steady state solution topologies described in the previous Sections
need not be the final solution of a set of time dependent equations. Depending
on the stability of the solutions, the flow may or may not settle on the steady 
state solution. However, it so happens that except for the solutions in regions
$NSA$ and $NSW$, numerical results actually match with analytical
results. The accuracy of the matching primarily
depends on the ability of the numerical code to conserve angular momentum
and energy. In Chakrabarti \& Molteni (1993) one dimensional simulations
and in Molteni, Lanzafame \& Chakrabarti (1994) and Molteni, Ryu
\& Chakrabarti (1996) two dimensional simulations have been
presented using parameters from $SA$ and $SW$ regions. 
Simulation by diverse codes produced similar results.
These suggest that these analytical solutions could 
be used for rigorous tests of the codes in curved spacetime. The parameters 
from regions $NSA$ and $NSW$ show unstable behaviours in a multidimensional
flow (Chakrabarti et al, 1996, in preparation) though in
strictly one dimension they match with the analytical work. Preliminary 
solutions of cold flows (${\cal E}=1$)
have already been reported in the literature (Ryu, Chakrabarti \& Molteni, 1996). 

Fig. 6 shows the Mach number and density variations in an one dimensional
flow around a Schwarzschild black hole (Molteni, Ryu and Chakrabarti,
1996). The solution is chosen from region $SA$ so that 
analytically a stable shock is expected in a thin
flow of polytropic index $\gamma=4/3$. The solid curve shows the
analytical solution: at the shock the density goes up as matter
virtually stops behind the shock and the Mach number goes down
from supersonic to subsonic. The long-dashed and the short-dashed
curves are the results of simulations using Total Variation Diminishing 
(TVD) method (see, Harten 1983; Ryu et al, 1995) and the Smoothed 
Particle Hydrodynamics (SPH) method (see, Monaghan, 1985; 
Molteni \& Sponholz, 1994). The results show that the shocks form
in an accretion flow, and they are stable. These simulations 
also verify the shock free solutions.

\noindent{\bf Spectral Properties of Generalized Accretion Disks}

How does an accretion disk radiate? What are the observational
signatures of a black hole? How to distinguish a black hole and a neutron
star from observations? In order to answer these questions
one clearly needs to solve the hydrodynamic equations 
in conjunction with radiative transfer. In the first approximation one
could construct a realistic accretion disk model based on
analytical solutions mentioned above, and then compute the
radiations emitted out of this model flow.

In Fig. 3, we showed that above a critical viscosity or below a critical
angular momentum or inner sonic point location, the flow
does not have a shock wave  after deviating from a Keplerian disk. Since with 
height the sonic point is expected to be closer to a black hole, and
a Keplerian disk above a sub-Keplerian flow may be Rayleigh-Taylor unstable,
it may be worthwhile to consider the important alternative that the
flow viscosity decreases with height. This makes a solution topology of the 
kind as in Fig. 3b to be on the equatorial plane with a Keplerian
disk close to a black hole, while a solution 
with a shock (Fig. 3a) at a higher elevation. Fig. 7 shows a 
typical disk model where Keplerian disk is flanked by 
sub-Keplerian `halo' which undergoes a shock wave at around $r\sim 10-30$ 
in Schwarzschild geometry and roughly at half as distant in extreme Kerr 
geometry (Chakrabarti \& Titarchuk, 1995).
Photons from the Keplerian disks are in the optical-UV range in
the case of massive black holes at galactic centers, and in the
soft X-ray range in the case of small black holes in X-ray binaries.
These photons are intercepted by the post-shock flow and are
energized through inverse Compton process, and are eventually
re-radiated at higher energies (as soft X-rays in massive
black holes, or, in hard X-rays in smaller black holes). 
First consider a galactic black hole candidate of mass $\sim 5M_\odot$.
When accretion rate of the Keplerian component is very low, the
soft X-ray from the Keplerian component is weak, but 
the hard X-ray intensity from the post-shock region is very strong.
This is known as the hard state. As the accretion rate increases, electrons
in the post-shock region become cooler and the object eventually 
goes to the soft state where only the soft-X rays from the
Keplerian disk dominates. Occasionally, one finds a weaker hard component tail
of slope $\alpha_\nu\sim 1.5$ even in the soft state. This is interpreted 
(Chakrabarti \& Titarchuk, 1995)
as due to the `bulk motion Comptonization' where the converging flow energizes
soft photons not due to thermal Comptonization (i.e., by thermal motion of 
the hot electrons) but due to direct transfer of bulk motion momentum 
of the cool electrons to cooler photons (Blandford \& Payne, 1981;
Titarchuk, Mastichiadis \& Kylafis, 1996).

Figs. 8(a-b) show the variation of intensity $L_\nu$ of
radiation as a function of frequency $\nu$ (Chakrabarti \& Titarchuk, 1995).
In Fig. 8a, we show contributions from various components of the disk.
The dotted curve represents the black body radiation from the optically
thick Keplerian flow. The long dashed curve represents the
fraction of these photons which were intercepted by the postshock
flow and were energized by hot electrons of the post-shock flow through 
thermal Comptonization processes. The dash-dotted curve is the reflection
of these hot photons from the Keplerian disk. The solid
curve gives the overall spectra which has a bump in the soft 
X-ray and a power law in the 2-50 keV region.
The above figure is for $M_{BH}=5M_\odot$ and 
dimensionless disk accretion rate of ${\dot m}_{\rm disk}=0.1 {\dot M}_{Edd}$.
In Fig. 8b, we show the computed spectra as the accretion rate of the
Keplerian component is varied. The solid, long-dashed, short-dashed 
and the dotted curves are for accretion rates ${\dot m}_{\rm disk}=
0.001,\ 0.01,\ 0.1,\ 1$ respectively. We also plot the 
dash-dotted curve which includes the effect of the bulk-motion 
Comptonization and shows a weak hard tail. The bulk motion Comptonization
is the process by which the cool electrons deposit their
momentum (while rushing towards the black hole horizon) onto photons
and energize them. The spectral index $\alpha_\nu$
($L_\nu \propto \nu^{-\alpha_\nu}$) of the weak hard tail
computed from Chakrabarti \& Titarchuk (1995) model is around $1.5$, 
and is often observed in black hole candidates in their soft states.
The conclusions regarding general spectral shape 
remain unchanged when computations are done for supermassive black hole.

\noindent{\bf Comparison With Observations and Concluding Remarks}

As an example of the success of our understanding of black hole
accretion processes, we use the aforementioned theoretical understanding
to fit a spectrum of GRS 1009-45 (X-ray Nova in Vela)
obtained by MIR-KVANT module experiments (Titarchuk et al. 1996)
The source was discovered by
WATCH/GRANAT all-sky monitor (Lapshov et al., 1993; IAUC 5864) and verified
by BATSE/GRO (Harmon et al., IAUC 5864).  
Fig. 9 shows the TTM and HEXE data (the photon flux versus energy;
Sunyaev et al. 1994) and the best theoretical fit given
by the considerations of the converging inflow in the post-shock
region. The parameters, namely, mass of the black hole (M), accretion rate
in Keplerian disk (${\dot m}_{disk}$), accretion rate in the sub-Keplerian
disk (${\dot m}_h$), the shock location ($R_{sh}$), the
post-shock temperature ($T_{sh}$), fraction of the soft photons
intercepted by the post-shock bulge ($H$) and the distance ($D$)
of the object, which are derived from the best-fit of
the data seem to be: 
$M = 3.7554 M_\odot$, ${\dot M}_{disk}/{\dot M}_{Edd}
= {\dot m}_{disk}=1.94$, ${\dot M}_h / {\dot M}_{Edd}= {\dot m}_h=
1.1$,  $R_{sh}=11.74R_g$, $T_{sh}=5.59 (KeV)$, $H=0.015$, $D=5.0 (Kpc)$
The $\chi^2=1.03$ for this fit. The general agreement clearly
vindicates the claim that a black hole accretes a significant
amount of sub-Keplerian matter. In a neutron star accretion, the
weak hard tail is not expected.

Some of the black hole candidates show quasi-periodic oscillations
of its spectra in some range of hard X-rays (Dotani, 1992).
Moteni, Sponholz, \& Chakrabarti (1996) and Ryu, Chakrabarti, Molteni
(1996) show that these oscillations could be due to dynamic oscillations 
of the standing shock waves or even the sub-Keplerian region itself. 
The frequency and amplitude modulation
of the radiation, as well as the time variation of
the frequencies match well with observations.

In this review, we showed that understanding accretion processes on a 
black hole must necessarily include the study of sub-Keplerian 
flows and how they combine with Keplerian disks farther away. 
We showed that the so called viscous transonic flows are stable, and 
constitute the most general form of accretion flows. We showed that the
accretion shock could play an important role in energetics of the
radiated photons. Indeed we showed that hard and soft states of black hole
candidates as well as Quasi-Periodic Oscillations could be
explained if shock waves were assumed to be present. Both of these
observations along with our model could be used to obtain the
mass and the accretion rates of black holes.  Success of these solutions
clearly depend on more accurate observations of steady state and time
varying spectra of the black hole candidates.

\newpage

\centerline{REFERENCES}

\noindent Anderson, M.R. \& Lemos, J.P.S, MNRAS, 233, 489.\\
Blandford, R.D. \& Payne, D. G., 1981, MNRAS, 194, 1033.\\
Bondi, H., 1952, MNRAS,  112, 195\\
Chakrabarti, S.K. 1989, ApJ, 347, 365 (C89)\\
Chakrabarti, S.K. 1990, Theory of Transonic Astrophysical Flows 
(Singapore: World Scientific, Singapore, 1990) (C90)\\
Chakrabarti, S.K. 1996a, Physics Reports, 266, No. 5-6, 238 \\
Chakrabarti, S. K., 1996b, ApJ, (June 20th issue).\\
Chakrabarti, S. K., 1996c, MNRAS, (in press).\\
Chakrabarti, S. K.  {\it M.N.R.A.S}, 1993, {\bf 261}, 625.\\
Chakrabarti, S.K., \&  Molteni, D. 1993, ApJ, 417, 671.\\
Chakrabarti, S. K. \& R. Khanna M.N.R.A.S, {\bf 256}, 300.\\
Chakrabarti, S. K., \& Titarchuk, L. G., 1995, ApJ, 455, 623.\\
Dotani, Y., 1992, in Frontiers in X-ray Astronomy (Tokyo: 
Universal Academic Press) 152.\\
Harten, A., 1983, J. Comput. Phys., 49, 357.\\
Harmon, B.A., Zhang, S.N., Fishman, G.J. \& Paciesas, W.S., 1993, IAU. Circ. No. 5864 \\
Kaniovsky, A., Borozdin, K., \& Sunyaev, R., 1993, IAU. Circ. No. 5878\\
Lapshov, I., Sazonov, S., \& Sunyaev, R., 1993, IAU Circ. No. 5864\\
Malkan, M. 1982, ApJ, 254, 22\\
Molteni, D., Lanzafame, G., \&  Chakrabarti, S.K. 1994, ApJ, 425, 161 (MLC94)\\
Molteni, D. \& Sponholz, H. 1994, in Journal of Italian
Astronomical Society, Vol. 65-N. 4-1994, Ed. G. Bodo \& J.C. Miller\\
Molteni, D., Sponholz, H., \&  Chakrabarti, S.K. 1996, ApJ, 457, 805\\
Molteni, D., Ryu, D., \&  Chakrabarti, S.K. 1996, ApJ (Oct. 10th, in press)\\
Monaghan J.J., Comp. Phys. Repts., 1985, 3, 71\\
Novikov, I., \& K.S. Thorne. 1973. in: Black Holes,
eds. C. DeWitt and B. DeWitt (Gordon and Breach, New York)\\
Ryu, D., Brown, G., Ostriker, J.P. \& Loeb, A., ApJ, 1995, 452, 364\\
Ryu, D., Chakrabarti, S.K. \& Molteni, D., ApJ (Submitted)\\
Shakura, N.I., \& Sunyaev, R.A. 1973, A\&A, 24, 337\\
Sun, W.H. \&  Malkan, M. 1989, ApJ, 346, 68\\
Sunyaev, R. A.  et al.\ 1994,  Astronomy Letters,  20, 777\\
Titarchuk, L., Mastichiadis, A. \& Kylafis, N. 1996, ApJ,  (submitted)\\
Titarchuk, L.G. et al. 1996 (in preparation).\\ 

\centerline{\bf Figure Captions}

{
\noindent  {\bf Fig. 1:} Classification of the parameter space (central box)
in the energy-angular momentum plane in terms of various
topology of the black hole accretion. Eight surrounding boxes 
show the solutions from each of the independent regions of the
parameter space. See text for details.

\noindent {\bf Fig. 2(a-b):} Mach number $M$ is plotted against logarithmic 
radial distance $r$. Contours are of constant entropy. $a=0.5$, $\lambda=3$.
In (a), ${\cal E}=1.003$ and in (b), ${\cal E}=1.007$. $O$ and $I$ denote
the outer and inner sonic points respectively. Single arrow shows the
accretion flow path through $O$ while double arrow traces
the path of more stable shocked flow which passes through both $I$ and $O$.

\noindent  {\bf Fig. 3:} Mach number variation (a-d) and angular
momentum distribution (e-h) of an isothermal viscous transonic flow.
Only the topology (a) allows a shock formation. Transition to
open (no-shock) topology is intiate by higher viscosity or lower angular
momentum or inner sonic point location. In (e-h), flow angular 
momentum (solid) is compared with Keplerian angular momentum (dotted).

\noindent  {\bf Fig. 4(a-b):} Ratio of (a) disk angular momentum to 
Keplerian angular momentum ($\lambda/\lambda_{Kep}$) and
(b) radial to azimuthal velocities ($v_r/v_\phi$) are shown
for a few solutions. Parameters are marked on the curves.  Note that $r_{Kep}$,
where the flow joins a Keplerian disk, depends inversely on the
viscosity parameter. 

\noindent {\bf Fig. 5:} Comparison of rotational viscous stress $\sigma_\phi^r|_{rot}$
(short dashed curves) with complete viscous stress $\sigma_{\phi+}^r$ (solid)
along the supersonic branch (passing through outer sonic point)
for (a) a prograde flow (upper panel) and  (b) a retrograde flow (lower
panel). Also shown is $d\Omega/dr$ (dotted curves). For 
comparison, results for the subsonic branch is also shown (long dashed).
Note the change in sign of the shear near the horizon. $\sigma_{\phi+}^r$ 
does not vanish on the horizon, but $\sigma_\phi^r|_{rot}$ and $\sigma_{\phi-}^r$ do.

\noindent {\bf Fig. 6:} Comparison of analytical and numerical results
in a one-dimensional accretion flow which allows a standing shock. The long
and short dashed curves are the results of the TVD and SPH simulations
respectively. The solid curve is the analytical result for the
same parameters. Upper panel is the mass density 
in arbitrary units and the lower panel is the Mach number of the flow.  

\noindent {\bf Fig. 7:} Schematic diagram of the accretion processes
around a black hole. An optically thick,
Keplerian disk which produces the soft component
is surrounded by an optically thin
sub-Keplerian halo which terminates in a standing shock
close to the black hole. The postshock flow Comptonizes soft photons
from the Keplerian disk and radiates them as the hard component. Iron line
features may originate in the rotating winds.

\noindent {\bf Fig. 8(a-b):} Analytical computation of the
emitted radiation from a black hole. (a)
Contributions of various components to the net
spectral shape (solid). Dotted, short dashed, long dashed and dash-dotted
curves are the contributions from the Shakura-Sunyaev disk $r>r_s$,
the reprocessed hard radiation by the Shakura-Sunyaev disk, reprocessed
soft-radiation by the postshock disk $r<r_s$ and the hard
radiation reflected from the Shakura-Sunyaev disk along the observer
($\mu\!=cos \theta =\!0.4$). Parameters are ${\dot m}_{disk}\!=\!0.1$,
${\dot m}_h\!=\!1$, and $M\!=\!5\ M_\odot$. (b)
Variation of the spectral shape as the accretion rate
of the disk is varied. ${\dot m}_{disk}\!=\!0.001$ (solid line),
$0.01$ (long-dashed line), $0.1$ (short-dashed line), and $1$ (dotted line).
The dash-dotted curve represents the hard component from convergent 
inflow near the black hole and has the characteristics of the slope 
$\sim 1.5$ in soft state.

\noindent {\bf Fig. 9:} Spectral fits of broad band X-rays from 
GRS1009-45 using our composite disk model. Data is obtained
from MIR-KVANT experiments as reported in Sunyaev et al. (1994).

\newpage
\noindent Appeared in the Proceedings of XVIIIth Conference Indian Association
for General Relativity and Gravitation (IMSc Report No. 117)
Feb. 1996, Eds. G. Date and B. Iyer

\noindent Authors's address AFTER November 26th, 1996:\\

\noindent Prof. S.K. Chakrabarti\\
\noindent S.N. Bose National Center for Basic Sciences\\
\noindent JD Block, Sector -III, Salt Lake\\
\noindent Calcutta 700091, INDIA\\

\noindent e-mail: chakraba@bose.ernet.in  OR chakraba@tifrc2.tifr.res.in \\

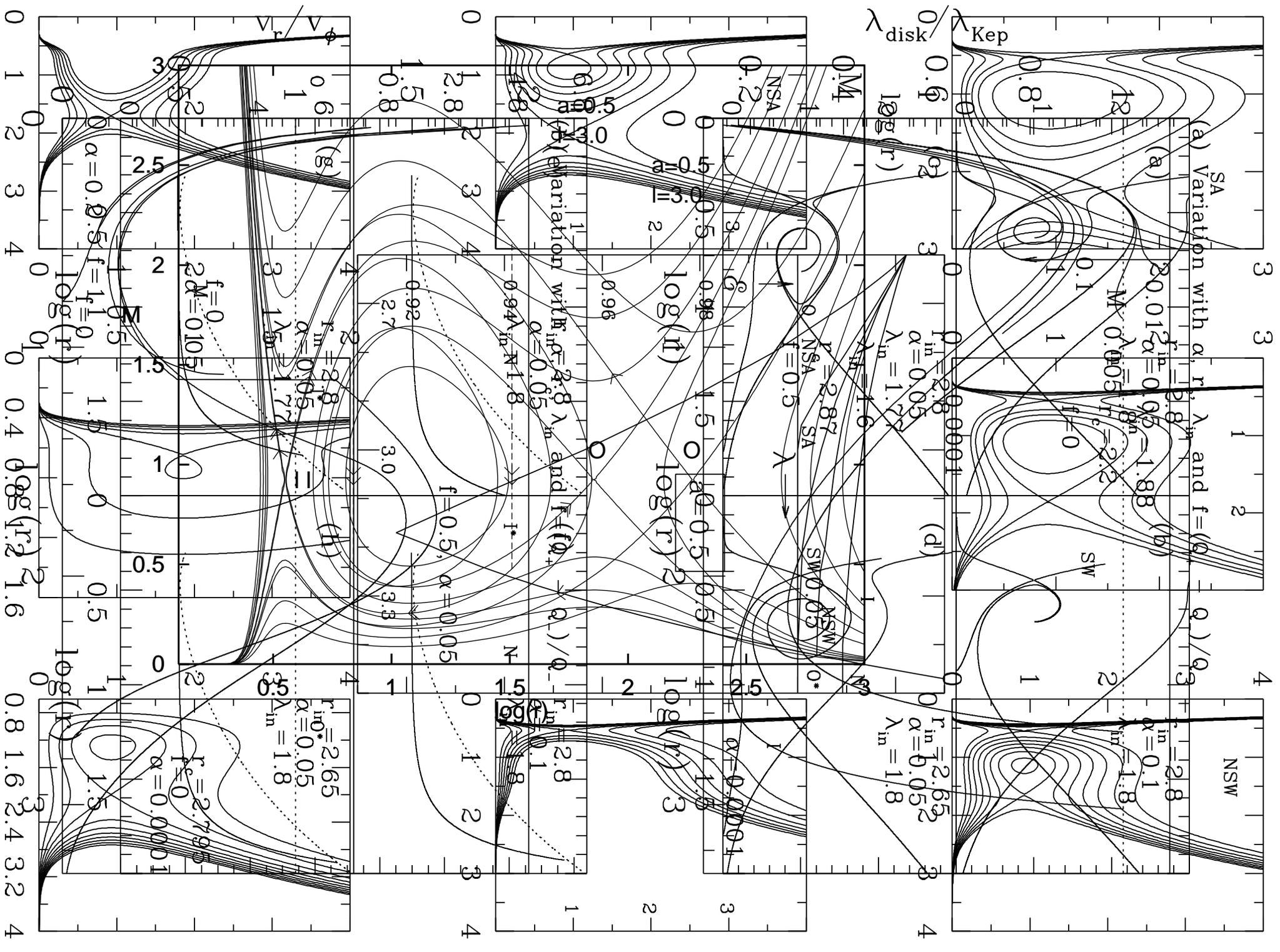
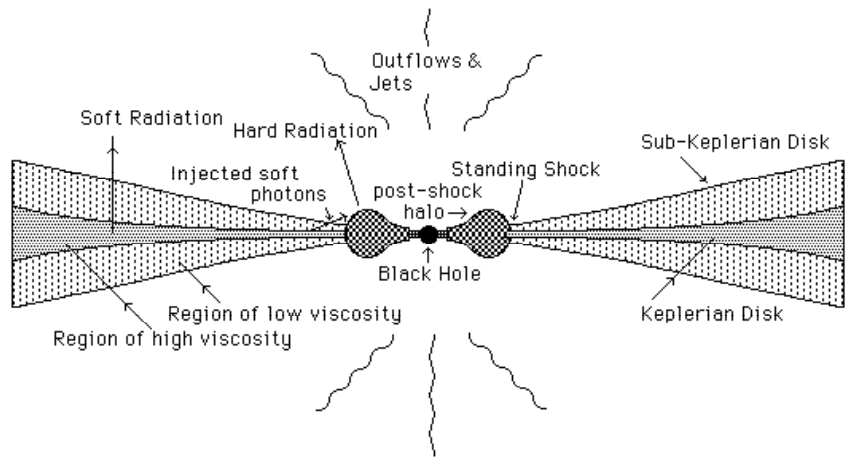
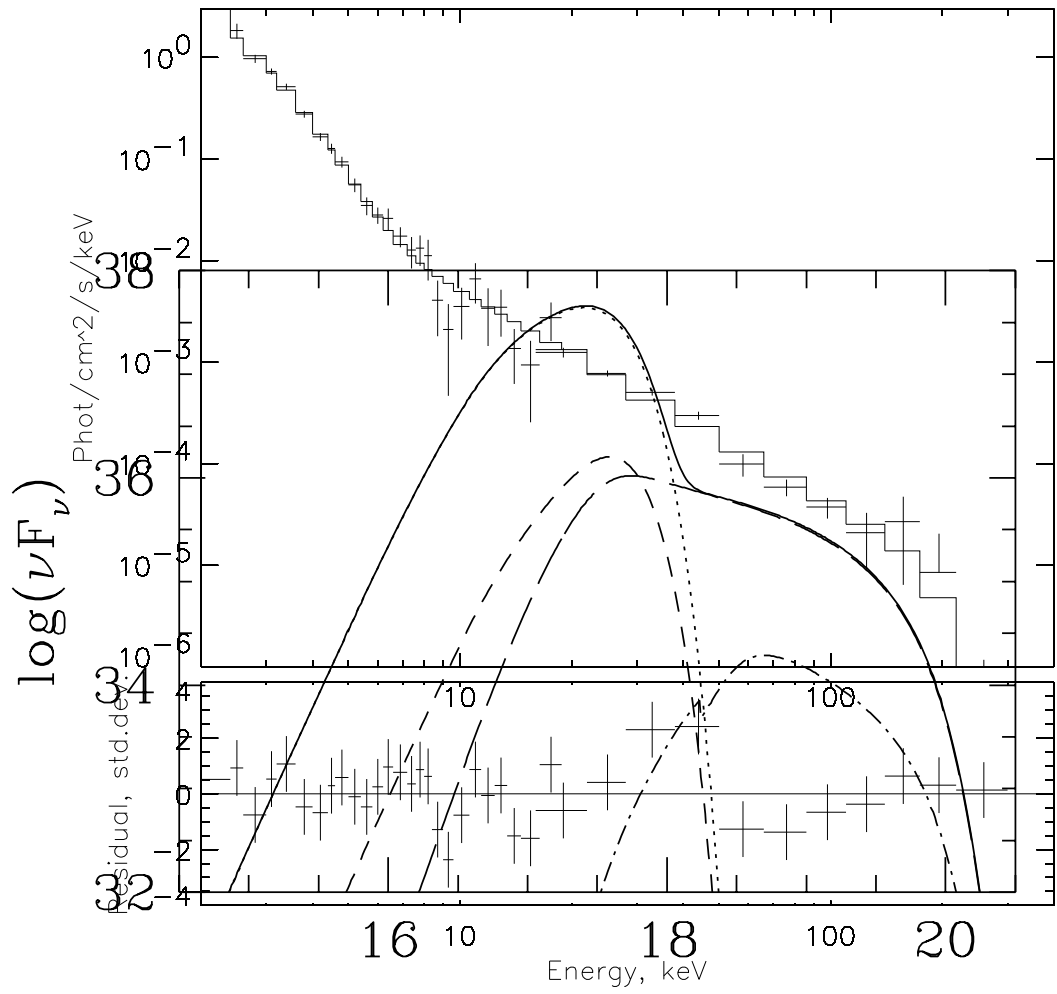
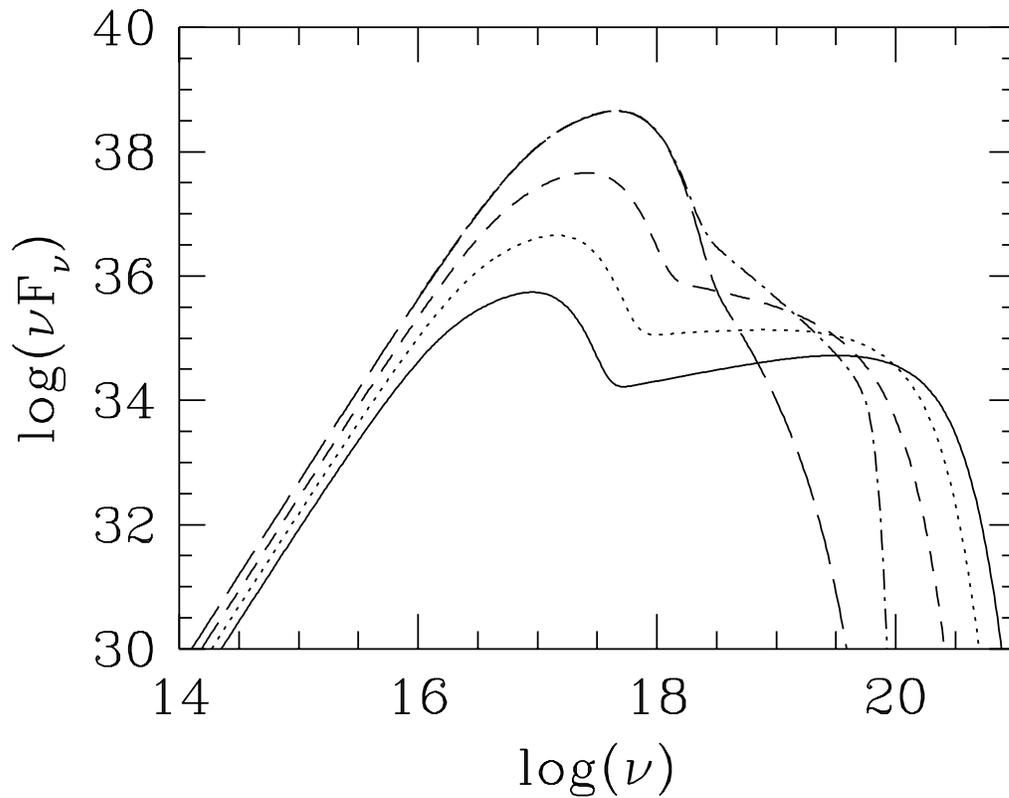
\end{document}